\documentclass[preprint,12pt]{elsarticle}

\usepackage{amssymb}

\usepackage{geometry}
\usepackage{caption}
\newcommand{\beginsupplement}{%
        \setcounter{table}{0}
        \renewcommand{\thetable}{S\arabic{table}}%
        \setcounter{figure}{0}
        \renewcommand{\thefigure}{S\arabic{figure}}%
     }
\makeatletter
\define@key{Gin}{resolution}{\pdfimageresolution=#1\relax}
\makeatother

\usepackage{wrapfig}

\journal{arXiv}

\begin{document}

\begin{frontmatter}



\title{Radiomic features of multi-parametric MRI present stable associations with analogous histological features in brain cancer patients}

 \author[label1]{Samuel Bobholz}
 \author[label2]{Allison Lowman}
 \author[label3]{Alexander Barrington}
 \author[label2]{Michael Brehler}
 \author[label1]{Sean McGarry}
 \author[label4]{Elizabeth J. Cochran}
 \author[label5]{Jennifer Connelly}
 \author[label6]{Wade M. Mueller}
 \author[label2]{Mohit Agarwal}
 \author[label2]{Darren O'Neill}
 \author[label7]{Anjishnu Banerjee}
 \author[label2,label3]{Peter S. LaViolette}
 \address[label1]{Department of Biophysics, Medical College of Wisconsin}
 \address[label2]{Department of Radiology, Medical College of Wisconsin}
 \address[label3]{Department of Biomedical Engineering, Medical College of Wisconsin}
 \address[label4]{Department of Pathology, Medical College of Wisconsin}
 \address[label5]{Department of Neurology, Medical College of Wisconsin}
 \address[label6]{Department of Neurosurgery, Medical College of Wisconsin}
 \address[label7]{Department of Biostatistics, Medical College of Wisconsin}

\begin{abstract}
MR-derived radiomic features have demonstrated substantial predictive utility in modeling different prognostic factors of glioblastomas and other brain cancers.  However, the biological relationship underpinning these predictive models has been largely unstudied, with the generalizability of these models also called into question.  Here, we examine the localized relationship between MR-derived radiomic features and histology-derived “histomic” features using a dataset of 16 brain cancer patients. Tile-based radiomics features were collected on T1W, post-contrast T1W, FLAIR, and DWI-derived ADC images acquired prior to patient death, with analogous histomic features collected for autopsy samples co-registered to the MRI.  Features were collected for each original image, as well as a 3D wavelet decomposition of each image, resulting in 837 features per MR image and histology image.  Correlative analyses were used to assess the degree of association between radiomic-histomic pairs for each MRI.  The influence of several confounds were also assessed using linear mixed effect models for the normalized radiomic-histomic distance, testing for main effects of scanners from different vendors and acquisition field strength.  Results as a whole were largely heterogenous, but several features demonstrated substantial associations with their histomic analogs, particularly those derived from the FLAIR and post-contrast T1W images.  These most-associated features typically presented as stable across confounding factors as well.  These data suggest that a subset of radiomic features are able to consistently capture texture information about the underlying tissue histology.
\end{abstract}

\begin{keyword}
Radiomics \sep MRI \sep Glioma \sep Brain Cancer


\end{keyword}

\end{frontmatter}


\section{Introduction}
\label{S:1}

Gliomas are an aggressive and often deadly form of primary intracranial tumor, representing 81 percent of malignant brain tumors (1). Current standard treatment includes surgical removal of the tumor followed by administration of radiation and chemotherapy (2,3).  Critical to maximizing the efficacy of these treatments is the detection of interval tumor progression and  monitoring areas of potential treatment effect using non-invasive imaging.  Multiparametric magnetic resonance imaging (MP-MRI) is central to diagnosing a glioma, monitoring its progression, and assessing the efficacy of treatment, as well as providing information regarding neuroanatomy and the physical properties of the tissues. Typical clinical protocols include pre and post-contrast T1-weighted images (T1, T1+C), and T2-weighted fluid attenuation inversion recovery (FLAIR) images, attempting to delineate enhancing, non-enhancing, and necrotic tumor components (4,5).  Diffusion-weighted images are also often collected in order to calculate the apparent diffusion coefficient (ADC) map, which is used prognostically to identify areas of restricted diffusion that may manifest as tumor recurrence (6–8) or stroke (9–11).  Despite their current utility, advancements in clinical imaging and tumor monitoring are necessary to more accurately identify the active, non-enhancing tumor components and inform improved treatment directions. 

Recently, MP-MRI has been used for a quantitative feature extraction process known as radiomics.  These radiomic features attempt to quantify aspects of an image such as intensity distributions, spatial relationships, and textural heterogeneity (12–15).  Feature extraction is typically calculated over a given region of interest (ROI), with shape-based features, histogram-based first order features, and a range of texture features across transformations of the original ROI matrix (16). Extensions of these pipelines have also calculated similar features for 3-D wavelet decompositions of the original image, which decompose the original image into its high and low frequency components and provide enriched edge-related information (17,18).  The end result of these image processing techniques are mineable datasets of quantitative information that can be compared to clinically relevant metrics. In studies of Glioblastoma Multiforme (GBM) and other brain cancers, several groups have demonstrated the utility of radiomic features in predicting prognostic factors such as survival time, anti-angiogenic treatment response, IDH1 mutation status, and differentiation of GBM from central nervous system lymphomas as well as lower grade astrocytomas (19–23). 

Despite the promising results shown with standard radiomics-based techniques, several methodological limitations have prevented their widespread adoption. Questions regarding the generalizability of radiomics-based predictive models have been raised by studies assessing the stability of the features across MR acquisition parameters (24–27). Current studies of radiomic features tend to focus on feature extractions across the whole tumor for subject-level classification, but studies of localized radiomic features have been limited.  Both the clinical need for better localization of active, non-enhancing tumor and the desire for biological validation of radiomics-based signatures further highlight this gap in the current literature.  An initial step in bridging this gap is to demonstrate localized associations between radiomic features and similar features of the underlying histology in order to demonstrate congruence between MR-derived features and features of the underlying tissue.  
	
	In order to address this initial step towards localized radiomics-based modeling, this study compares localized radiomic features to similarly calculated features of co-registered histology images, referred to here as “histomic” features.  This study uses MRI co-registered tissue samples from brain cancer patients acquired at autopsy to explore the association between tile-based radiomic features and their histomic analogs.  Specifically, this study tests the hypotheses that 1) radiomic features demonstrate substantial associative relationships with their histomic analogs, 2) the relationship between radiomic and histomic features is stable across scanner vendors and acquisition field strengths.

\section{Methods}
\label{S:2}

\begin{figure}[hbt!]
\vspace{0pt}
\centering\includegraphics[width=1\linewidth]{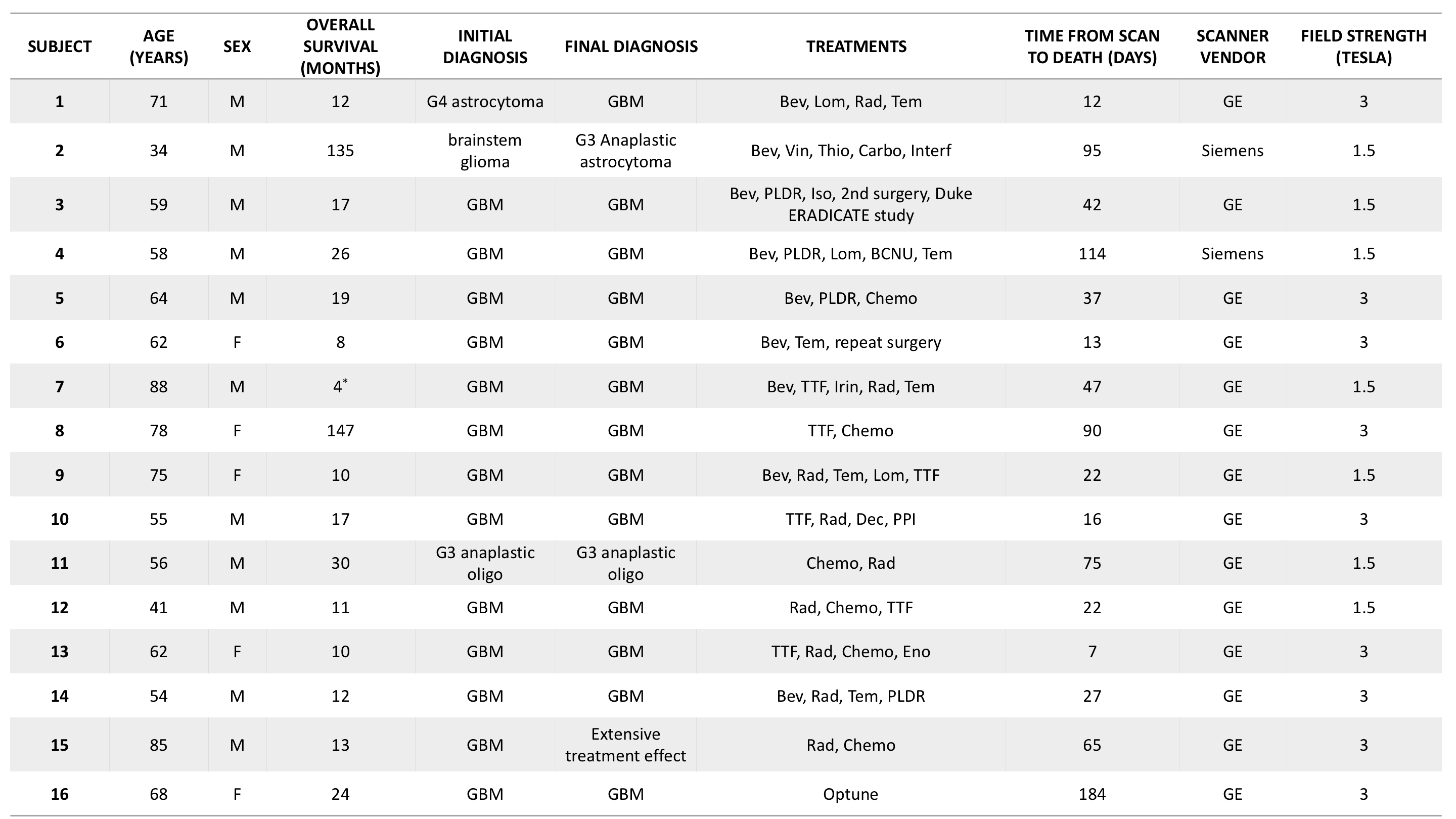}

\caption*{Table 1: Demographic and Clinical Characteristics of Sample.  Overall survival time is calculated from first surgery, except in case deonoted with *, in which case surgery was not performed and survival is calculated from first appearance on MRI.  Bev = bevacizumab, Lom = Lomustine, Tem = Temozolomide, Vin = Vincristine, Thio = Thio triethylene thiophosphoramide, Carb = carbamazepine, RNAI = RNA interference therapy, PLDR = pulsed low-dose rate radiotherapy, Iso = isotretinoin, Irin = irinotecan, Car = carmustine, Chemo = unspecified chemotherapy, Rad = unspecified radiation therapy, TTF = tumor treating fields, Dec = dexamethasone, PP1 = polyphyllin I, Eno = enoxaparin sodium.}

\end{figure}

\begin{figure}[hbt!]

\vspace{-20pt}
\centering\includegraphics[width=1\linewidth]{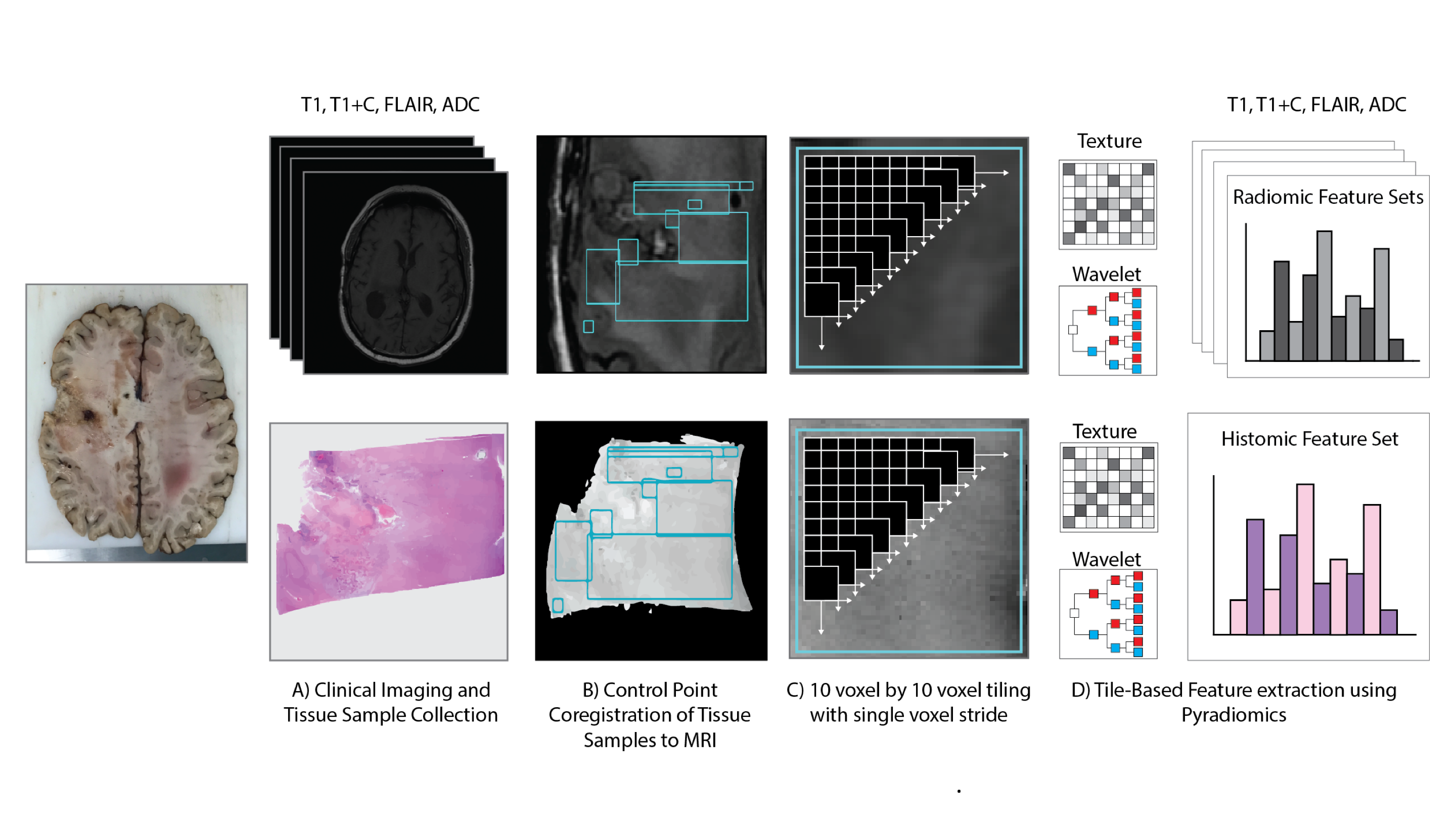}
\caption{Schematic representation of the data collection process.  For each subject, A) clinical MRI and HE-stained tissue samples of notable brain regions were collected, B) greyscale tissue samples were coregistered to the MRI using manual control point warping, C) tile masks were defined by using a 10 by 10 voxel frame with a single voxel stride across valid regions of the histology/MRI, and D) radiomic features were calculated across the tiles for each MR image and histomic features were calculated across the same tiles for the histology.}
\end{figure}

\subsection{Patient Population}

Sixteen patients with pathologically confirmed primary brain tumors were enrolled for the brain tissue component of this IRB approved study.  A brief outline of the clinical history of each subject is presented in Table 1 and a diagrammatic representation of the data collection process is presented in Figure 1.

\subsection{Image Acquisition and Preprocessing}

T1-weighted pre- and postcontrast images (T1 and T1+C, respectively), diffusion weighted images, and Fluid-attenuated inversion recovery (FLAIR) images were acquired from each subject prior to death for clinical purposes. Images were acquired on our institutional MRI scanners, including 1.5T and 3T GE (General Electric Health, Waukesha, Wisconsin) and Siemens (Siemens Healthineers, Erlangen, Germany) magnets.  Acquisition parameters for an example subject at 1.5T include (repetition time/echo time): T1 spin-echo sequence (T1), 666/14 ms; post contrast T1 spin-echo acquired with intravenous gadolinium (T1+C), 666/14 ms; apparent diffusion coefficient (ADC), calculated from diffusion-weighted images (DWI) acquired with an inversion recovery sequence using B=0 sec/mm2 and b=1000 sec/mm2, 10,000/90.7 ms; and FLAIR, acquired with an inversion recovery sequence, 10,000/151.8 ms and TI of 2,200 ms, all images acquired with submillimeter in-plane resolution.  

All MR images for a given patient were rigidly aligned with the FLAIR image using FSL’s FLIRT tool (28–30). After co-registration, T1, T1+C, and FLAIR images were intensity normalized by dividing voxel intensity by its standard deviation across the whole brain (31).  Images were visually inspected before and after processing to ensure images were free of any artifacts that may confound analyses.

\subsection{Ex-vivo Histology Processing}

Upon death, the brains of each subject were removed during autopsy and placed into 10 percent buffered formalin within a 3D-printed cage based on the subject’s most recent MRI to maintain structural integrity with respect to the imaging data during fixation, as previously published (32).  After approximately 2 weeks, the brain samples were sliced using slicing molds printed to delineate the axial slices from the most recent MRI.  Tissue samples were collected from each subject based on enhancement on imaging, suspected presence of tumor, or pathologist-determined diagnostic relevance.  The samples were then processed, embedded in paraffin, cut, and stained with hematoxylin and eosin (HE).  The full slides were photographed at 40X magnification using an Olympus sliding stage microscope.  Matlab 2018b (MathWorks Inc. Natick, MA) was then used to process each individual tile, followed by down-sampling and compiling of all other tiles from the sample to generate a single image for each slide.  Upon examination of these images, samples from each subject showed representative regions of both tumor and non-tumor characteristics, allowing assessment of the radiomic-histomic relationship across a range of tissue pathologies.  

\subsection{MRI-Histology Co-Registration}

Previously published in-house software (Hist2MRI Toolbox, written in Matlab) was used to precisely align histology images to the MRI (19,32–35).  Manually defined control points were applied to align each composite histology image with the analogous anatomical features in the FLAIR sequence MRI.  Digital photographs were taken at the time of the brain cutting and sample collection in order to precisely define the location of the histological sample with respect to the MRI.  Samples were spatially aligned to the MRI slice that best represented the sample’s location by visually inspecting photographs of the brain slices acquired before and after each sample was collected (19,32–35).  

\subsection{Feature Extraction}

For each grayscale, MR-space histology image, regions of interest (ROI) were manually drawn to designate areas of the image with valid histological information (i.e. free from tears, folds). These ROIs were then used to generate tiles for use in radiomic feature extraction using a 10 voxel by 10 voxel sliding frame.  Single-voxel strides in each dimension were used to define tiles across the ROIs (n=54,067), which were used as localized masks for calculation of the radiomic and histomic feature sets.  

Pyradiomics v2.1.1 was used generate the radiomic and histomic feature sets for each tile (16,18).  First order features (FO, n=18), gray level co-occurrence matrix (GLCM, n=24), gray level dependence matrix (GLDM, n=14), gray level run length matrix (GLRLM, n=16), gray level size zone matrix (GLSZM, n=16), and neighboring gray tone difference matrix (NGTDM, n=5) features were calculated for each MR image.  The same features were additionally calculated on eight 3D wavelet decomposition (3DWD) images of each MRI, generated by applying all combinations of high and low pass filters in each dimension.  This resulted in a total of 837 radiomic features per MRI modality.  The same 837 features were then calculated for the grayscale histology image at MR-resolution for each tile, resulting in an analogous histomic feature set.  

\subsection{Statistical Analyses}
\subsubsection{Experiment 1: Correlative Analyses}

In order to examine direct monotonic associations between analogous radiomic-histomic feature pairs, Spearman’s rank correlations were computed between the radiomic feature from each MRI and its histomic analog.  Peripheral analyses revealed that time between MR acquisition and death influenced the radiomic-histomic relationship (Supplemental Figure 1); thus, results are reported both before and after correction for this effect using partial Spearman correlations. Due to the large sample size of this analyses relative to the overall subject count, p-values were not considered a valuable indicator of meaningful effects; thus, effect sizes are reported here.

\subsubsection{Experiment 2: Stability Analyses}

Stability analyses were conducted in order to observe the effects of various confounding on the generalizability of radiomic-histomic relationships.  The tile-wise log-Euclidean-distance (TLED) between the overall radiomic feature set of each MRI and the overall histomic feature set was used to assess global differences in radiomic-histomic associativity.  The TLED probability density function for GE and Siemens scans collected at 3T were plotted to assess global effects of vendor, and the probability density function for 1.5T and 3T GE scans were plotted to assess global effects of field strength.  Differences in distributions were quantified using the Kolmogorov-Smirnoff test.  

In addition, feature-wise assessments of vendor and field strength effects were performed using linear mixed effect models.  Separate models were fit to assess these effects on the normalized difference between each analogous radiomic-histomic pair.  Two models were fit for each pair: 1) an assessment of vendor (GE vs. Siemens) as a main effect amongst subjects with MRI acquired at 3T and 2) an assessment of field strength (1.5T vs. 3T) as a main effect amongst subjects with MRI acquired using a GE scanner.  Both models for each feature pair included time between MRI and death as a covariate, as well as random effects for subject and tissue sample.

\begin{figure}[hbt!]

\vspace{0pt}
\centering\includegraphics[width=1\linewidth]{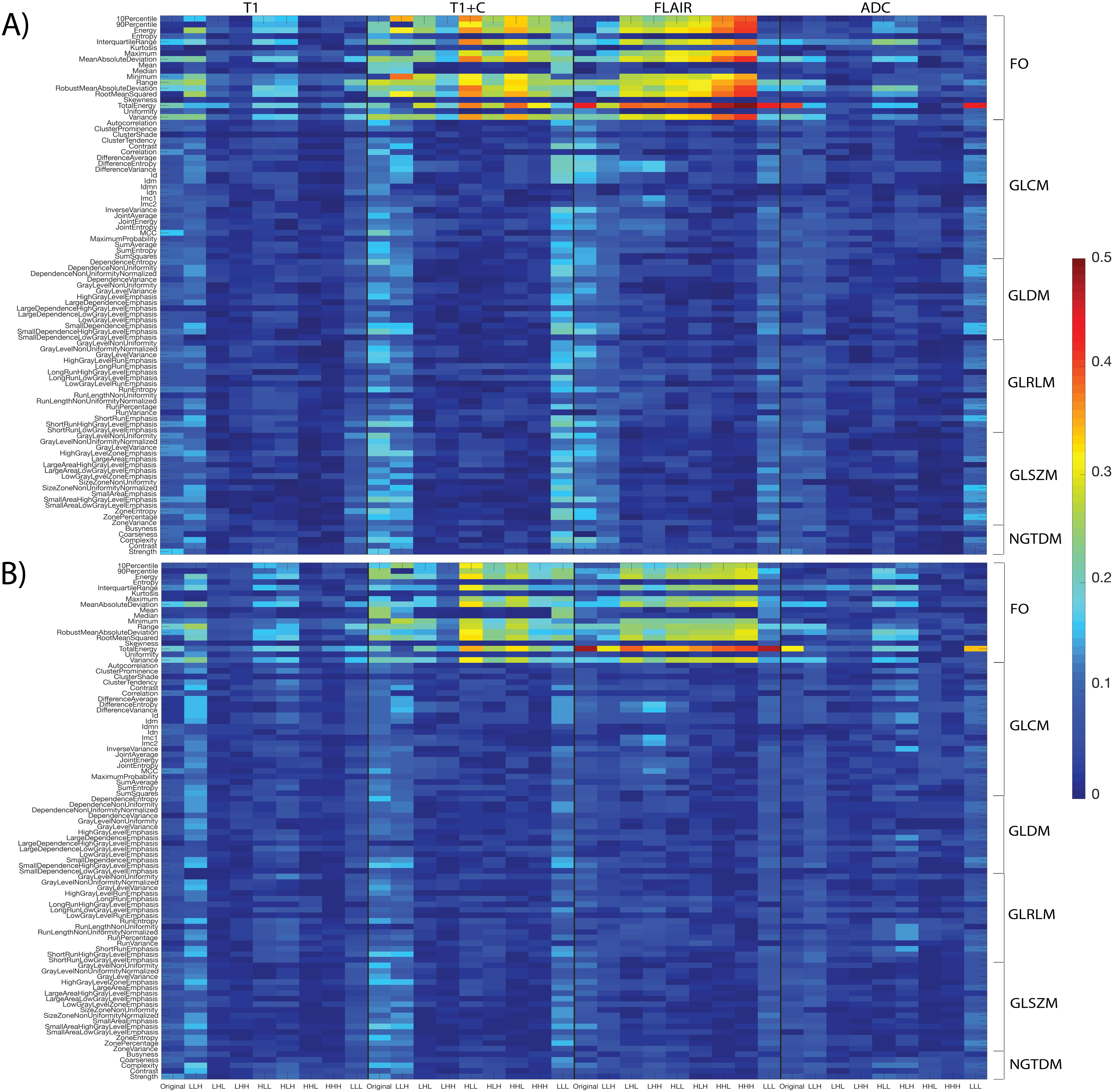}
\caption{Heatmap of Spearman’s Rho values between analogous radiomic-histomic feature pairs, presented by feature A) before and B) after correction for time between MRI and death. Note, first order features in general showed greater radiomic-histomic associations than the other five categorical feature sets.}
\vspace{0pt}
\end{figure}

\section{Results}

\subsection{Experiment 1: Correlative Analyses}

Figure 2 shows the resulting correlation heatmaps, displaying the Spearman’s Rho between the radiomic and histomic version of each feature.  FO features tended to show the strongest radiomic-histomic association, with very few GLCM, GLDM, GLRLM, GLSZM, and NGTDM features demonstrating meaningful radiomic-histomic relationships.  Across the 3DWD images, the strength of association between FO features tended to increase, whereas higher order associations tended to dissipate across the wavelet decomposition images.  Radiomic-histomic associations sorted by strength are presented in Figure 3 in order to compare feature strengths across image types, with T1+C and FLAIR images showing the strongest radiomic-histomic relationships.  Overall, controlling for time between MR acquisition and death tended to decrease the strength of radiomic-histomic associations, but did not affect the general trends regarding image types and feature sets seen in the uncorrected data.  

\begin{figure}[hbt!]

\vspace{0pt}
\centering\includegraphics[width=1\linewidth]{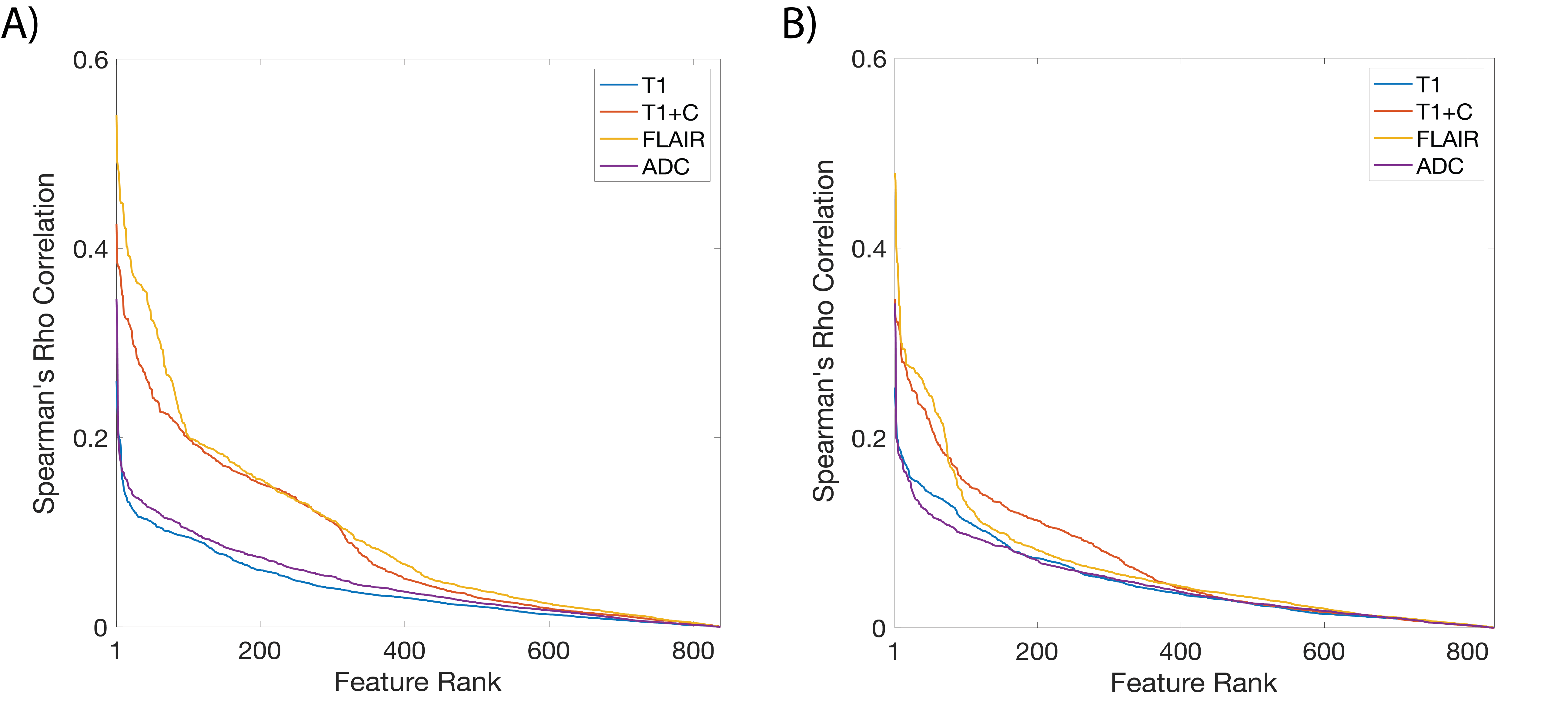}
\caption{Ranked feature associations for each MR image contrast A) before and B) after correction for time between MRI and death. The FLAIR image generally shows the strongest associations, closely followed by the T1+C image.  }
\vspace{0pt}
\end{figure}

\subsection{Experiment 2: Stability Analyses}

TLED distributions for each confound analyses are presented in Figure 4.  MR scanner vendor tended to have a larger influence on the overall TLED than field strength, though most scans displayed minor to moderate influence of both confounds on the overall TLED.  The feature-wise assessment of these confounds is presented in Figure 5, which plots the standardized mixed model coefficient of the confounds against the radiomic-histomic association for each feature.  These results confirm that differences in scanner vendor tend to outweigh differences in acquisition field strength.  These plots also indicate that the features showing the strongest radiomic-histomic relationships tend to have less substantial confounding effects across most image modalities.

\section{Discussion}

This study explored the histological underpinnings of tile-based MP-MRI radiomic features in GBM and other brain cancer patients.  Several radiomic features demonstrated substantial associations with their histomic analogs, suggesting that these aspects of the MRI directly characterize the same features of the underlying tissue histology.  These features are shown to be relatively robust across different confounding factors, such as MR scanner vendor and acquisition field strength.  These findings, taken as a whole, begin to provide a neuroanatomical context for radiomics-based models of brain cancer characteristics.

To the best of our knowledge, this study is the first assessment of localized relationships between radiomic features and analogous histomic features calculated on histology samples.  Our findings suggest that radiomic features are able to capture localized information about the underlying histology of the tissue, motivating a thrust towards radiomics-based models for localized tissue information.  Past studies attempting to map localized tissue information using MRI have focused on the use of biopsy cores as the source of ground truth (37–40).  While this allows for characterization of tumor in the MRI-enhancing region, the use of autopsy samples in this study enabled an assessment of the radiomic-histomic relationship in both tumor and non-tumor tissue.  Whole brain generalizability will improve the clinical utility of MRI-based tissue feature maps.  Thus, this study motivates future investigations using autopsy samples as ground truth in order to capture whole-brain heterogeneity of tissue features.

\begin{figure}[hbt!]
\vspace{0pt}
\centering\includegraphics[width=1\linewidth]{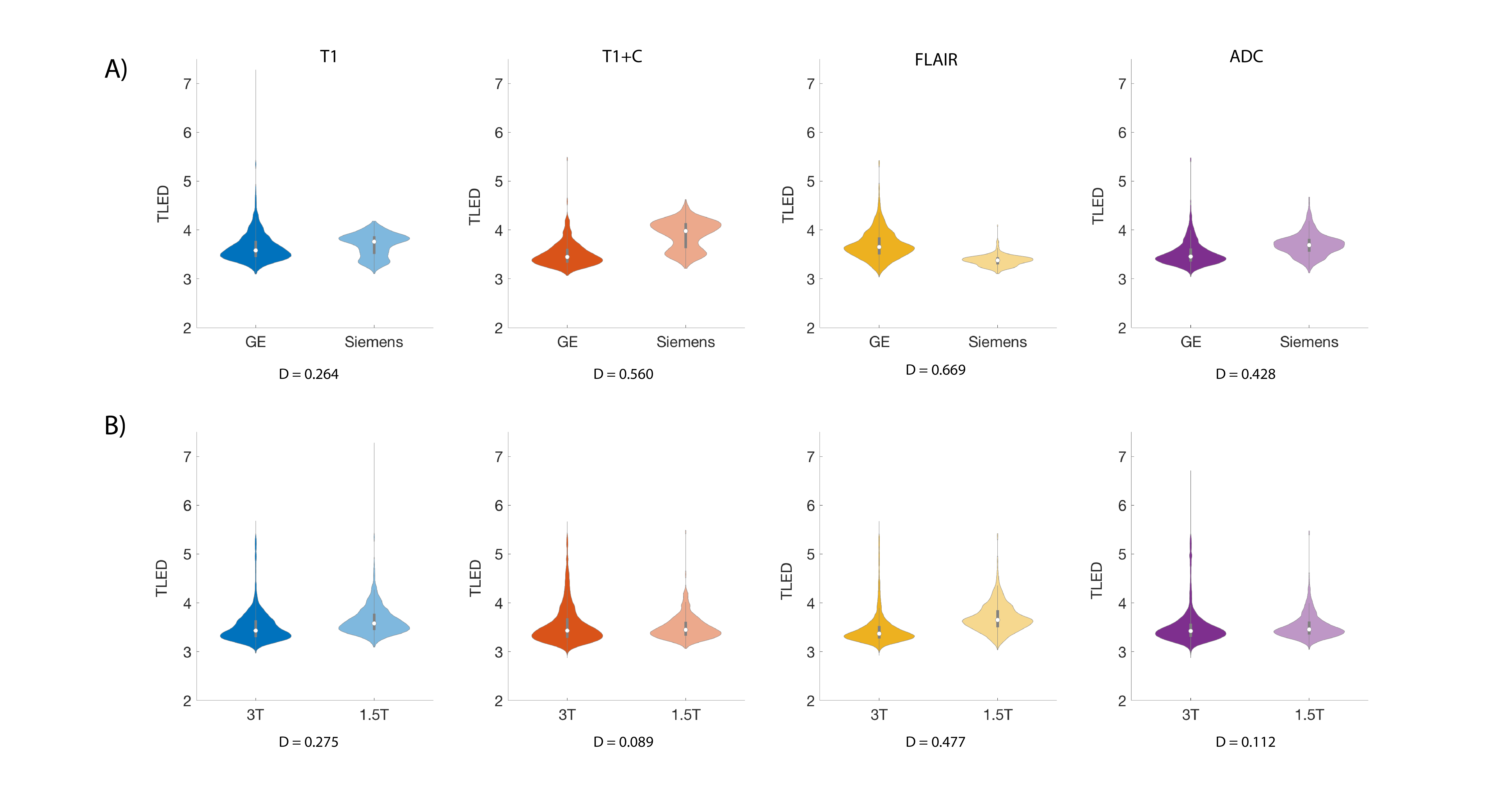}
\caption{Distributions of Tile-wise Log Euclidean Distance (TLED) between radiomic and histomic features, grouped by A) scanner vendor, and B) acquisition field strength.  Differences in distributions, quantified by K-S statistic D, indicate different radiomic-histomic similarity distributions between classes.  Both factors show modest to substantial influences on the overall radiomic-histomic relationship, though these results suggest scanner vendor may be a stronger overall confound than acquisition field strength.}
\vspace{0pt}
\end{figure}

Despite the benefits of measuring the radiomic-histomic relationship across autopsy samples, the time between the MRI scan and tissue fixation at death becomes a new confounding factor, as subtle changes in the disease state may occur during this time.  When assessing radiomic-histomic relationships after controlling for the duration of this period, nearly all features show mild to moderate decreases in associative strength.  Despite these effects, the overall strength and patterns of association remained largely intact, suggesting that minor statistical artifacts in the uncorrected data did not falsely manifest these general trends.  Larger studies are thus warranted to better assess the validity of this claim, as well as to provide better characterizations of MRI-to-fixation duration effects on different MRI-to-tissue mappings.  

The most notable contrasts within the associative analyses (Figure 2) include the general strength of FO features compared to various higher order features, as well as the modulation of this relationship across the 3DWT.  Generally, FO features presented with more substantial associations than those of higher order features, though higher order features of the T1+C and FLAIR images still showed some weak associations.  This split in associative strength was compounded by the 3DWT results, with FO features increasing in associative strength and higher order features decreasing in strength across most wavelet decompositions.  The lower frequency wavelet decompositions (LLL and LLH) deviated from this pattern and showed some increase in higher-order associations.  Future research into the inter-feature relationships will provide insight as to whether or not these wavelet decompositions provide new information or simply preserve higher order associations found in the original image.

\begin{figure}[hbt!]
\vspace{0pt}
\centering\includegraphics[width=1\linewidth]{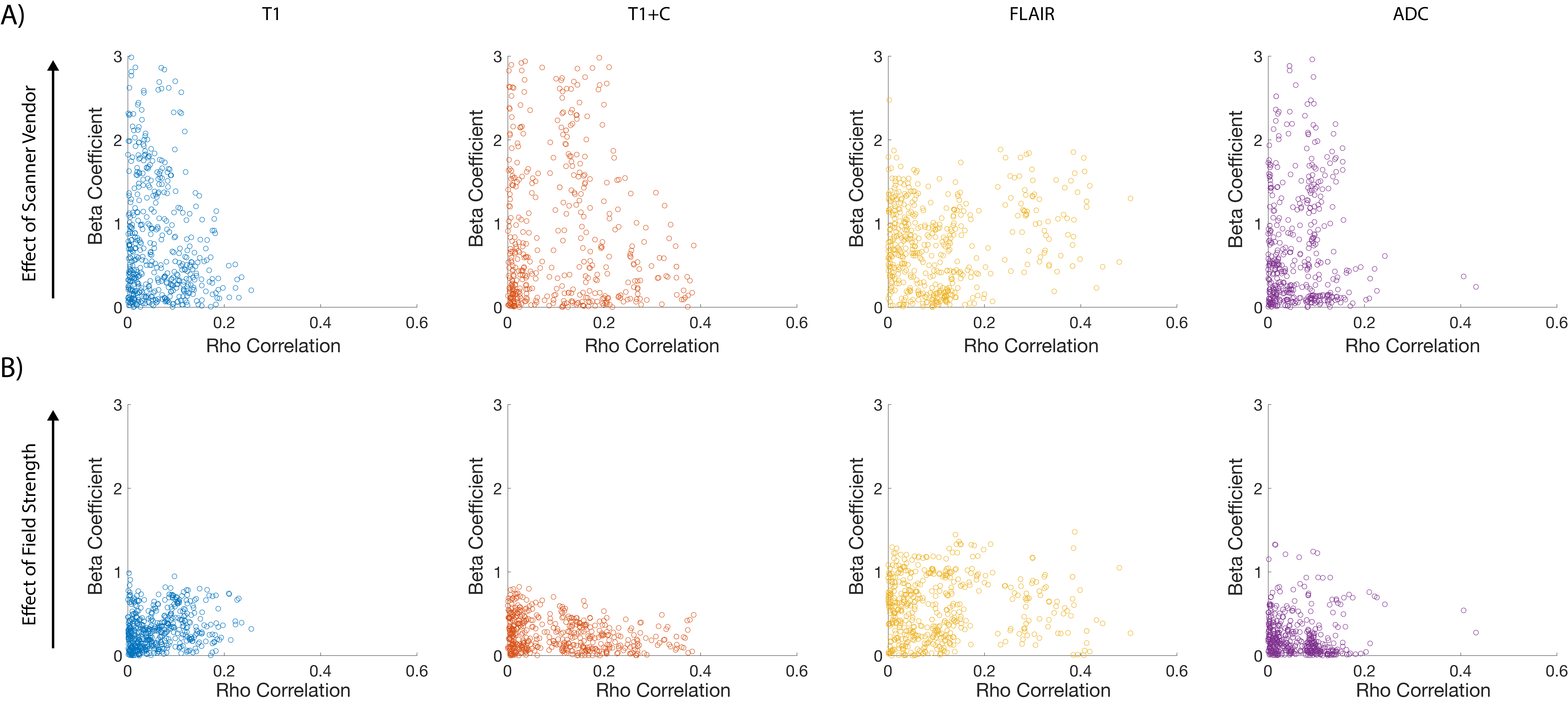}
\caption{Effects of A) scanner vendor and B) acquisition field strength presented by feature.  Each plot shows the standardized beta coefficient from each mixed model analyses, plotted against the radiomic-histomic association for each feature.  Aside from the FLAIR	image, featues with the highest radiomic-histomic associations tended to show the lowest influences of scanner vendor and acquisition field strength.}
\vspace{0pt}
\end{figure}

Comparing across scan types, the FLAIR image tended to have the strongest associations with histology texture (Figure 3).  Given the immense diagnostic utility of FLAIR images in delineating different tumor regions, this does not come as a great surprise, and supports our use of the FLAIR image as the optimal reference image for histology co-registration.  The T1+C image also demonstrated substantial radiomic-histomic associations, which suggests that the addition of contrast agent enhances the degree of texture preservation seen in the MRI.  The features of the T1+C image with the strongest radiomic-histomic relationships also tended to be more stable across scanner vendor and acquisition field strength than those of the FLAIR.  These results suggest that the contrast agent may provide new information relevant to tissue texture beyond that of the standard T1 image, which showed substantially weaker radiomic-histomic associations.

The weak radiomic-histomic associations observed in the T1 and ADC image relative to the FLAIR and T1+C images imply that these MRI modalities are not as representative of underlying tissue histology texture.  This does not necessarily discount the predictive utility of these images as much as it highlights what sort of information these modalities may provide.  Whereas the magnitude of the ADC image in a given region may provide robust and insightful information regarding some diagnostic criteria, the features of the ADC and T1 images may not be as relevant to providing histological texture information.  This distinction is important as studies move towards assessing the biological underpinnings of radiomics.  The lack of texture preservation seen in these results point future studies towards non-textural candidates for biological validation of radiomic features derived from these modalities. 

A clinical degree of MR scanner and field strength heterogeneity was present in the MR acquisitions of this sample, which was used to examine the influence of these factors on the radiomic-histomic relationship.  Previous studies have demonstrated several radiomic features susceptible to differences across different scanners (41–43); this study adds to this literature by providing a breakdown of how well features represent histological texture across scanner vendor and field strength.  As desired, the features with the greatest degree of association between radiomic and histomic analogs tended to have the least substantial effects of scanner and field strength, with the notable exception of the evenly-confounded FLAIR features.  Statistical artifacts of the confounds may influence the weaker associations observed in this data set, but generally these results point to a subset of features characterized by both strong and stable radiomic-histomic associations.  Larger follow-up studies assessing radiomic-histomic relationships may be able to reveal additional features that accurately characterize tissue texture for particular scanner vendors and acquisition parameters.  

This study is not without its limitations, however.  The relatively small sample size of this study calls for replication of these results in larger data sets to confirm the generalizability of the patterns seen here.  This study was also limited to primary brain cancer patients, which limits the scope of the radiomic-histomic relationship these results imply.  Future studies of radiomic-histomic relationships in other diseases will be an important step in establishing more general cases for the relationship between MRI and tissue texture.  The use of a tile-based prediction to study localized information does not allow for the use of shape- and size-based radiomic features, which often provide useful weights in radiomics-based modeling of prognostic factors.  Though this study statistically controlled for the duration between MRI acquisition and death, it is possible that these effects are not adequately addressed as covariates in this small sample.  Future, large-scale studies of autopsy data will be essential to characterize the magnitude of this effect, as well as to address optimal strategies to account for these effects.  

In conclusion, this study provides a novel characterization of the radiomic-histomic relationship in an attempt to provide a neuroanatomical basis for radiomics analyses.  These results show a substantial degree of heterogeneity in the strength and stability of radiomic-histomic relationships but reveal a subset of radiomic features that stably reflect information about the underlying histology.  This study provides the groundwork for future investigations into the quantitative pathological validation of radiomics analyses as currently performed, as well as underscores an opportunity for radiomics-based predictions of localized histological features with diagnostic and prognostic utility.

\section{Acknowledgements}

We would like to thank our patients for their participation in this study and our funding sources: American Brain Tumor Association DG14004, R01CA218144, R01CA218144-02S1, and R21CA23189201.

\section{Disclosures}

The authors of this manuscript have no disclosures to report.












\beginsupplement

\vspace{350pt}
\centering\includegraphics[width=1\linewidth]{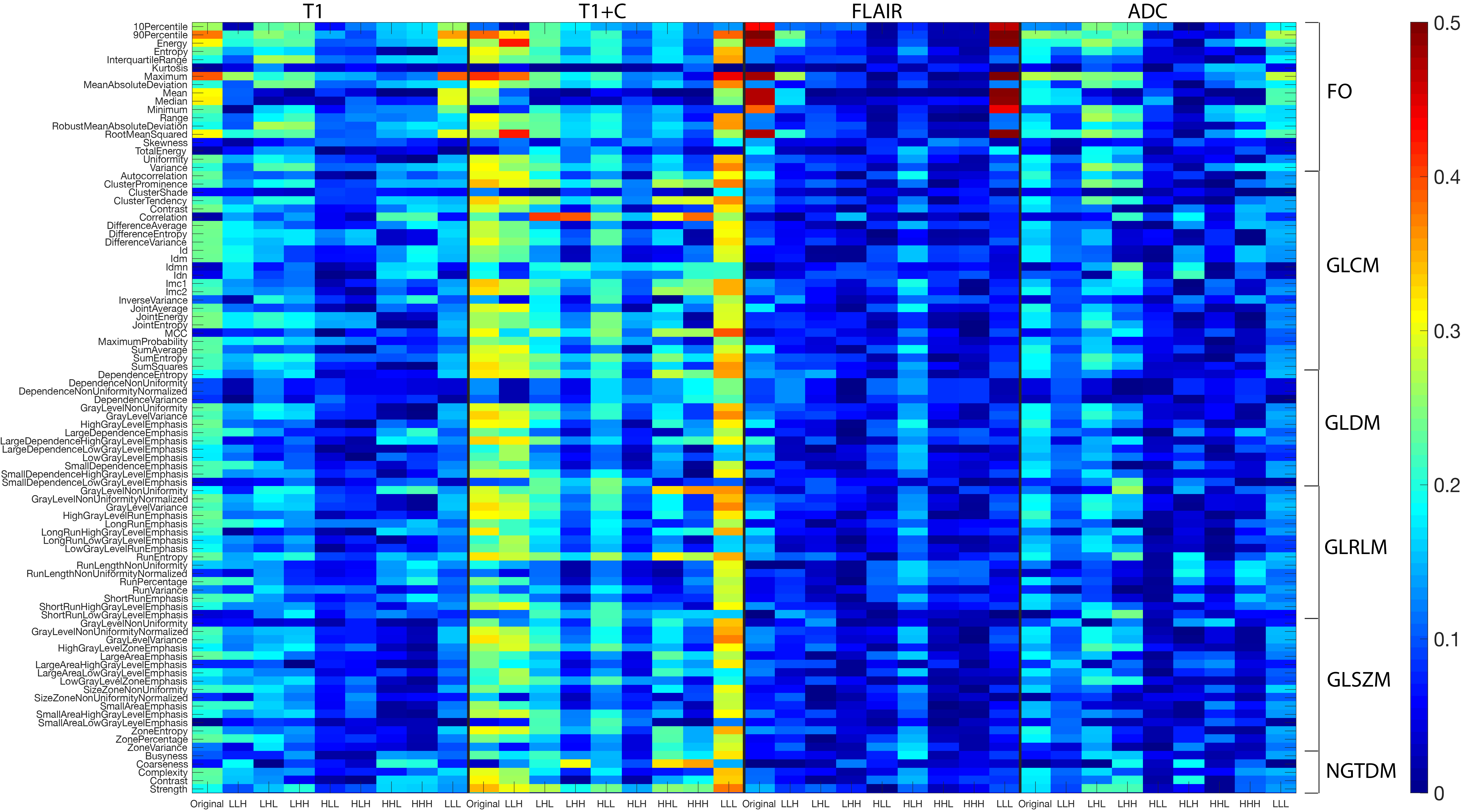}

\begin{flushleft}

Supplemental Figure 1: Spearman's Rho associations between the normalized radiomic-histomic feature difference and time between MRI and death, presented by feature.

\end{flushleft}

\vspace{0pt}

\end{document}